\documentstyle[12pt,equation]{article}
 1
\begin{document}

\title{\bf QUASI-ELASTIC SCATTERING IN THE INCLUSIVE ($^3$He, t)
REACTION }

\author{Neelima G. Kelkar \\
INFN--Laboratori Nazionali di Frascati,\\
 P.O. Box 13, I-00044
Frascati(Roma), Italy. \\\\
B.K.Jain\\
Nuclear Physics Division, Bhabha Atomic Research Centre,\\
Bombay-400085, India.}
\maketitle
\begin{abstract}
The triton energy spectra of the charge-exchange $^{12}$C($^3$He,t)
reaction at
2 GeV beam energy are analyzed in the
quasi-elastic nucleon knock-out region. Considering that this region is
mainly populated by the charge-exchange of a proton in $^3$He with a
neutron in the target nucleus and the final proton going in the
continuum, the cross-sections are written
in the
distorted-wave impulse approximation. The t-matrix for the
elementary exchange
process is constructed in the DWBA, using
one pion- plus  rho-exchange potential for the spin-isospin nucleon-
nucleon potential. This t-matrix reproduces the experimental data on the
elementary  pn $\rightarrow $np process. The calculated
cross-sections for the $^{12}$C($^3$He,t)
reaction at $2^o$ to $7^o$
triton emission angle are compared with the corresponding
experimental data, and are found in reasonable overall accord.
\end{abstract}
PACs numbers: 13.75.-n, 25.55.-e
\newpage
\section {\bf Introduction}
Because of the easy transferability
of sufficient energy in the intermediate energy nuclear collisions
to the nucleus, and the reduced effect
of the Pauli blocking, the quasi-free scattering forms a
major portion of the reactive content
in nuclear cross-sections at these energies. In experiments, this
fact is manifested by a broad
bump in the ejectile energy spectrum around
$\omega = q^2/2m^*$, where ($\omega, \mbox{\bf q}$) is the four
momentum transfer
to the nucleus ($m^*$ being the effective mass of the nucleon). The
width of this bump is correlated to the momentum spread of the nucleon
in the nucleus. In earlier times,
this aspect  was exploited much to gather directly the information
about
the single particle aspect, in particular the
shell model, of the nucleus through the study of the inclusive
(p,p$'$), (e,e$'$) reactions, and the exclusive (p,2p),
 (e,e$'$p) [1], and other reactions of similar type.
 In recent years, however, the focus on similar studies
 has shifted to the charge-exchange reactions, like (p,n), ($^3$He,t)
 in the quasi-free region [2,3]. This has happened because of
the discovery of strong Gamow-Teller excitations in
these reactions and a rather simple (Born term) description of the
 spin-isospin piece of the N-N interaction in terms of a one-
 pion- and a rho-exchange interaction [4]. It is felt that the
 study of these reactions in the quasi-free
 region, like the earlier quasi-free knock-out studies, would provide
 an
opportunity to explore the single particle spin-isospin  response
of the nucleus. This response, due to pi- and a rho-meson-exchange,
contains longitudinal as well as the transverse
components. Going beyond the single particle aspect, one expects that
these studies may also explore the particle-hole correlations
in the quasi-free region.
Theorists predict [5] that the particle-hole correlations, apart from
modifying the magnitudes, shift
the longitudinal response towards the lower
excitation energy and the transverse response towards
the higher excitation energy. In addition, it is also known that
these correlations also
renormalize the propagation of pions in the
nuclear medium.
Because of this, the study of the spin-isospin nuclear response to
various external probes has been a topic of great interest over the
past decade.
An extensive experimental study of the ($^3$He,t) reaction has been
carried out at Saturne [6], and  the (p,n)
reaction  at Los Alamos [2].
Theoretically too, several efforts have been made to study these
reactions. Alberico  et
al. [7] have developed a random phase approximatiom theory (RPA) of
the spin-isospin nuclear surface response and studied the contrast
between the
spin-longitudinal ($R_L$) and spin-transverse ($R_T$) part of the
nuclear response.
Their predictions are for nuclear matter and use a $(\pi +
\rho + g')$ model for the interaction. Ichimura et al. [8] have
improved upon this method and have calculated $R_L$ and $R_T$ by
the continuum RPA with
the orthogonality condition. They treat the nucleus as of
finite size
and present the cross-sections for $^{40}$Ca(p,p$'$) reaction at
E$_p$ = 500 MeV using the distorted wave impulse approximation.
However, notwithstanding these efforts, Bertsch et al. [9], while
discussing a number of experiments in a recent
critical review of this field, find that the effect of the residual
particle-hole correlations seen in the experiments in the quasi-free
region is much smaller than expected.

Considering the above observation of Bertsch et al. [9] as an
indication
of the weakness of the correlations (whatever may be the reason),
in the present paper we study the quasi-elastic peak region as being
populated by the charge-exchange knock-out of a neutron in the
target nucleus. The motivation for this work is to explore the
extent upto which the
experimental data could be accounted by the independent particle
framework alone. We have done the calculations in the DWIA, where
the interaction of the mass 3 particles in the continuum with the
nucleus is incorporated through the use of distorted waves.
As a typical case, we analyse the data  on the
$^{12}$C($^3$He,t) reaction at 2 GeV beam energy [3]. Specifically, we
assume that the
quasi-free region in this reaction is populated by the $^{12}$C($^3$He,tp)
reaction, where the proton in the final state arises due to the
charge exchange of a proton in $^3$He with a neutron in
1s or 1p shell in $^{12}$C.  Since
the experimental data for the $^{12}$C($^3$He,t) reaction in the
quasi-free region are of inclusive type, we do not include, in
our calculations, the
distortion potential for the proton in the final state. The main
effect of distortion of the proton in the $^{12}$C($^3$He,tp)
reaction is to remove the proton flux from this channel to other
channels, which in the inclusive measurements is included in the data.
Furthermore, since, due to strong absorption of the projectile
and ejectile, the charge-exchange between the $^3$He and $^{12}$C
nucleons occurs in the low density surface region of the nucleus,
we consider the elementary process, pn$\rightarrow$np, in the nucleus
as a quasi-free process. In the present case, of course,
this process is off-shell.
We construct the t-matrix for it following our earlier studies on the
elementary processes, p(n,p)n and p(p,$\Delta ^{++}$)n [10].
In this work, the t-matrix is constructed in the DWBA, using one-
pion plus rho-exchange potential for the $V _{\sigma \tau}$. This t-
matrix reproduces the experimental data on the p(n,p)n reaction. For the
p(p,$\Delta ^{++}$)n reaction, incidentally, in the same work it was
found that only
one-pion-exchange results agree with the experiments.

In section 2 we give the formalism for the ($^3$He,t) reaction. The
transition amplitude is written in a distorted wave impulse
approximation, as
mentioned above, and distortions of $^3$He and triton are treated
in the eikonal approximation. We also present briefly the procedure
to calculate the elementary t-matrix, t$_{\sigma \tau}$.

In the charge-exchange reaction, besides t$_{\sigma \tau}$, the
cross-section also receives contribution from the
iso-spin term, t$_\tau$. Since, in Boson-exchange models, this
t-matrix gets constructed from second- and higher-order Born terms only,
we have not constructed it here. We have used for it the
phenomenologically determined t-matrix of Franey and Love [11]. In
any case, as we shall see later, the contribution of t$_\tau$
to the ($^3$He,t) cross-section is not much.

The experimental data for the $^{12}$C($^3$He,t) reaction, as obtained
by Bergqvist et al. [3], exist for the triton energy spectrum
at  $2^o$ to
$7^o$ emission angles at 2 GeV beam energy. These spectra are inclusive.
The broad structure seen  in them between 1.9 and 2 GeV triton energy can
be ascribed to the quasi-free charge-exchange reaction. The theoretical
cross-sections corresponding to these spectra are obtained by first
calculating the double differential cross-section
$d^2\sigma /d\mbox{\bf $k_t$} d\mbox{\bf $k_p$}$, and then integrating
it over
the allowed kinematics of the outgoing protons and summing over the various
neutron states in the target nucleus. In section 3 we present
the calculated differential cross-sections. The calculations are done
with and without the $\rho$-exchange contribution in the interaction,
and compared with the experimentally measured spectra. We find a
reasonable overall agreement between the calculated and measured
cross-sections with pi- plus rho-exchange interaction.

\section{Formalism}

The differential cross-section for the triton energy spectrum in the
reaction $^3$He + A $\rightarrow $t + p + B is given in the lab. as,
\begin{equation}
\frac{d^2\sigma}{ dE_t d\Omega _t} =
\int d(cos\theta _p)
\times P \times
\langle |T_{BA}|^2 \rangle,
\end{equation}
where $\langle |T_{BA}|^2\rangle$ is the transition amplitude summed and
averaged over the spins in the initial and final states respectively.
The reaction mechanism is shown in fig.1. $\mbox{\bf k}_{He}$ is taken along
the z-axis and the x-z plane is defined by the vectors $\mbox{\bf
k}_{He}$ and $\mbox{\bf k}_t$. The factor
P in the above eq.  is given as,
\begin{eqnarray} \nonumber
&P& = \int \frac {d\phi_p}{ 2(2\pi )^5} \times  \\
& &\frac {k_t k^2_p m_{He} m_t m_p
m_B}{k_{He}[k_p(E_i-E_t)-k_{He}cos\theta _pE_p + k_tE_p
cos(\theta _{pt})]},
\end{eqnarray}
where $E_x$, $k_x$ and $m_x$ represent the energy, momentum and mass
respectively of the particle $x$. $\theta _{pt}$ is the emission angle
of the proton relative to the triton. For a given beam energy and
fixed value of
the triton four momentum, $k_p$ is determined by solving the appropriate
energy-momentum
conservation relations. The cross-section
$d^2\sigma / dE_t d\Omega _t$ is calculated by integrating over
all possible emission directions, $\theta _p$
and $\phi _p$, of the outgoing proton.

\subsection{Evaluation of $T_{BA}$ }
The transition amplitude for the reaction A($^3$He,t)B, in DWIA, is
given as,
\begin{equation}
T_{BA} =( \chi ^-_{\mbox{\boldmath $k$}_t},\mbox{\boldmath $k$}_p\langle B,
t,p|\sum_{ij} [t_{\sigma \tau}
(i,j) + t_\tau (i,j)]|
 A,^3He\rangle \chi ^+_{\mbox{\boldmath $k$}_{He}}),
\end{equation}
where $j$ represents the active nucleons in the target nucleus and
$i$ those in
$^3$He. $\chi _{He}$ and $\chi _t$ are the distorted waves for helium and
triton respectively. For protons, because of the inclusive nature of
the reaction, we use plane waves. This is appropriate, because,
as discussed in
the literature [12], the main effect of the distortion at intermediate
energies is absorptive. This results in the transfer of flux from
the given channel to other channels. In an inclusive reaction,
these channels are included in the measured cross-sections.

$t_{\sigma \tau}(i,j)$ is the spin-isospin t-matrix, and contains
the longitudinal and transverse components. It is off-shell.
In terms of its
central, $t^C$, and non-central, $t^{NC}$, components, we can write
it as
\begin{equation}
t_{\sigma \tau}(i,j)= \biggl [ t_{\sigma \tau}^C(\epsilon,\mbox{\boldmath $q$})
\mbox{\boldmath $\sigma$}_i \cdot \mbox{\boldmath $\sigma$ }_j +
t_{\sigma \tau}^{NC}(\epsilon,\mbox{\boldmath $q$}) S_{ij}(\hat{q})\biggr ]
\mbox{\boldmath $\tau$ }_i \cdot
\mbox{\boldmath $\tau$}_j,
\end{equation}
where $\mbox{\bf q}$ is the momentum transfer in the
reaction and $\epsilon$ is the energy at which the elementary t-matrix
for pn$\rightarrow$np needs to be evaluated. Actual evaluation of
this t-matrix
is described further below.

The tensor operator, $S_{ij}(\hat {q})$,
is defined as :
\begin{eqnarray} \nonumber
&S_{ij}(\hat {q})& = 3 \mbox{\boldmath $\sigma$}_i \cdot \hat {q}
\mbox{\boldmath $\sigma$}_j \cdot \hat {q} - \mbox{\boldmath $\sigma$}_i
\cdot \mbox{\boldmath $\sigma$}_j \\
& &= [\frac  {24 \pi}{5}]
^{1/2}\sum _M \sum_{\mu \nu} (-1)^{\mu +\nu} \sigma _{\mu} (i)
\sigma _\nu (j) \langle 1 1 -\mu -\nu | 2 M\rangle Y_{2M}(\hat {q})
\end{eqnarray}

To evaluate T$_{BA}$, we first observe that, around the
energy of interest of the continuum particles here
($\approx$2 GeV), the main effect of distortion is absorptive. The
dispersive effects are small. Therefore, in evaluating the elementary
t-matrix, $t_{\sigma \tau}$ (or $t_ \tau$), we approximate the
momentum transfer, $\mbox{\bf q}$, by that corresponding to the
asymptotic momenta of $^3$He and triton. The
$\langle |T_{BA}|^2 \rangle$ then factorizes as
\begin{equation}
\langle |T_{BA}|^2 \rangle=
\langle |G|^2\rangle  | \rho (\mbox{\boldmath $q$})|^2,
\end{equation}
where $\mbox{\bf q}=\mbox{\bf k}_{He}-\mbox{\bf k}_t$.
$\rho (\mbox{\bf q})$ is the spatial $^3$He$\rightarrow$t transition density
factor and is
normalized such that $\rho$(0) = 3. The factor
$\langle |G|^2\rangle $, after taking the expectation value of the
elementary t-matrix
over the spin-isospin wave functions of $^3$He and triton, and summing
and averaging the square appropriately over the spin projections
of $^3$He and triton, works out as,
\begin{eqnarray} \nonumber
&\langle |G|^2\rangle =&\frac
{4}{9}\frac {1}{(2J_B + 1)} \sum _{m_p M_B}\biggl [
 \sum _{m=-1}^{m=+1}[| t_{\sigma \tau}^C(\epsilon,\mbox{\boldmath $q$})
F^{-m,+1}(\mbox{\boldmath $Q$}) \\
& &+[\frac {24 \pi }{5}]^{1/2}t_{\sigma \tau}^{NC}(\epsilon,\mbox{\boldmath
$q$})
\sum _{\nu M} (-1)^{\nu} \langle 1 1 -m -\nu | 2 M \rangle
Y_{2M}(\hat {q})F^{\nu ,+1}(\mbox{\boldmath $Q$})|^2]\nonumber \\
& &+|t_{\tau}(\epsilon,\mbox{\boldmath $q$}) |^2 | F^{+1}( \mbox{\boldmath
$Q$})|^2\biggr ].
\end{eqnarray}
Here, $\mbox{\bf Q} =\mbox{\bf k}_{He} - \mbox{\bf k}_t -\mbox{\bf k}_p$ is
the momentum of the recoiling nucleus
in lab. In the impulse approximation, this momentum equals (with opposite
sign) to that of the struck
neutron in the target nucleus.
$ F^{\mu ,+1}(\mbox{\bf Q})$ is the `distorted' Fourier transform
of the spin-isospin overlap integral of the target and residual nucleus.
In configuration space, it is given by,
\begin{eqnarray}\nonumber
&F^{\mu ,+1}(\mbox{\boldmath $Q$})& = \langle B | \sum _i \chi ^{-*}
_{\mbox{\boldmath $k$}_t}(\mbox{\boldmath $r$}_i) \chi ^{- *}_{\mbox{\boldmath
$k$}_p}(\mbox{\boldmath $r$}_i)
 \chi ^+
_{\mbox{\boldmath $k$}_{He}}(\mbox{\boldmath $r$}_i) \sigma _{\mu}(i)
\tau _{+1} (i) | A \rangle \\
& &=\int d\mbox{\boldmath $r$} \chi ^{-*}
_{\mbox{\boldmath $k$}_t}(\mbox{\boldmath $r$}) \chi ^{- *}_{\mbox{\boldmath
$k$}_p}(\mbox{\boldmath $r$})
\chi ^+_{\mbox{\boldmath $k$}_{He}}(\mbox{\boldmath $r$}) \phi _{\mu,+1}^
{BA}(\mbox{\boldmath $r$}),
\end{eqnarray}
where $\phi _{\mu,+1}^{BA}(\mbox{\boldmath $r$})$ is the overlap integral
and is defined as
\begin{equation}
\phi _{\mu,+1}^{BA}(\mbox{\boldmath $r$})=\langle B|\sum _i \delta
(\mbox{\boldmath $r$}-\mbox{\boldmath $r$}_i)
\sigma _\mu (i)\tau _{+1}|A\rangle.
\end{equation}

For a shell model and a closed shell target nucleus (i.e. J$_A$=0), it
is easy to work out this integral. In this case,
for $F^{\mu ,+1}(\mbox{\bf Q})$ we eventually get,
\begin{eqnarray}\nonumber
&F^{\mu ,+1} (\mbox{\boldmath $Q$}) =&\sqrt 6 \sum _{m_s} \sum _{l m_l}
\langle l, 1/2, m_l, m_s |J_B,-M_B \rangle
\langle 1, 1/2, \mu, m_s |1/2, m_p \rangle  \\
& &\langle \chi _t^-\chi _p^- |\chi ^+_{He}, \phi _{l m_l} \rangle.
\end{eqnarray}

For the purely isospin-dependent term, in a similar way, we obtain,
\begin{equation}
\frac {1}{ (2 J_B + 1)} \sum _{M_B m_p} |F^{+1}(\mbox{\boldmath $Q$})|^2 =
\sum _{l,m_l}\frac {1}{(2l +1)} |\langle \chi _t^- \chi _p^- |
\chi ^+_{He} \phi_{l m_l} \rangle |^2.
\end{equation}

$\phi_{l m_l}$ in above eqs. is the spatial part of the wave function
in $nl$ shell of a neutron in
the target nucleus. It is normalized such that, $\langle \phi_{l m_l}
|\phi_{l m_l}\rangle=N_{nl}$, where $N_{nl}$ is the number of
neutrons in the shell.

In eq. (7) one may notice that the contributions of
the spin-isospin dependent
and the only isospin dependent part of the interaction to the
cross-section enter incoherently.
 The
central and non-central parts in the spin-isospin interaction, however,
add coherently.

\subsection{p(n,p)n t-matrix}
For the construction of this t-matrix we follow our earlier work [10].
In this work, including the effect of elastic and
other channels on $a\rightarrow b$ transition
in nucleon-nucleon scattering, at intermediate energies we write
\begin{equation}
t_{ba}(\mbox{\boldmath $k$}_i , \mbox{\boldmath $k$}_f ) =  (\chi
^{-*}_{\mbox{\boldmath $k$}_f},\langle b|
V_{\sigma \tau}|a \rangle, \chi ^+_{\mbox{\boldmath $k$}_i}),
\end{equation}
where $\chi $'s are the distorted waves for the pn relative motion.
They are the solutions of
potentials which describe the pn elastic scattering.
Below the pion threshold, these potentials are available from boson-
exchange models. However, in the energy region, which is of relevance
in the present work and is above the pion threshold,
these potentials need major modifications. In the absence of a
reliable estimate of such modifications, we have used
the eikonal
approximation (which is valid at higher energies) and have written
$\chi$'s directly in terms of the elementary elastic scattering
amplitude, f(k,q), as (for details see ref.[13]),

\begin{equation}
\chi ^+_{\mbox{\boldmath $k$}} (\mbox{\boldmath $r$}) = e^{i\mbox{\boldmath
$k$}
\cdot \mbox{\boldmath $r$}}\biggl [ 1+\frac {i}{k}
\int _0^{\infty }q
dq J_o(qb) f(k,q)\biggr ].
\end{equation}

Here the amplitude f(k,q) peaks at zero degree and falls off rapidly.
Near the
forward direction it can be reasonably parametrized as [14],
\begin{equation}
f(k,q) = f(k,0) exp(-\frac {1}{2} \alpha q^2),
\end{equation}
where, using the optical theorem, we can further write,
\begin{equation}
f(k,q) =[\frac {k}{4\pi }]\sigma _T(k)
(i + \beta (k)) exp(-\alpha (k)q^2 /2).
\end{equation}
Here, $\sigma _T$ is the total cross-section, $\beta $ is
the ratio of the real to imaginary part of the scattering
amplitude and $\alpha $ is the slope parameter in the pn scattering.
The values of these parameters depend upon the energy, k, of the pn
system.

The t-matrix, with the above parametrization, works out to be
\begin{equation}
t_{ba}(\mbox{\boldmath $k$}_i , \mbox{\boldmath $k$}_f ) =  \int
d\mbox{\boldmath $r$} e^{i\mbox{\boldmath $q$} \cdot \mbox{\boldmath $r$}}
exp [i\xi (b)]\langle b| V_{\sigma \tau}|a \rangle,
\end{equation}
where $\xi (b)$ is the phase-shift function, and is defined as
\begin{equation}
exp\biggl [ i \xi (b)\biggr ] = [1 - C exp(-b^2/2 \alpha )]+
iC\beta exp(-b^2 /2\alpha ),
\end{equation}
with $C = \sigma _T / 4 \pi \alpha $.

$V_{\sigma \tau}$ is the  spin-isospin
dependent transition potential. The major portion of this interaction,
as is well known [4],
arises from the one-pion- plus rho-exchange potential. We,
therefore, write
\begin{equation}
V_{\sigma \tau}(i,j) = \biggl [ V_{\pi}(t)
\mbox{\boldmath $\sigma$}_i \cdot \hat{q} \mbox{\boldmath $\sigma$}_j
\cdot \hat{q} + V_\rho (t)(\mbox{\boldmath $\sigma$}_i \times
\hat {q}) \cdot (\mbox{\boldmath $\sigma$}_j \times \hat {q})\biggr ]
\mbox{\boldmath $\tau$}_i \cdot \mbox{\boldmath $\tau$}_j,
\end{equation}
where
\begin{equation}
V_x(t) = -\frac {f_x^2}{ 3 m^2_x} F_x^2(t) \frac {q^2}
{ m_x^2 - t}.
\end{equation}
$t = \omega ^2 - \mbox{\bf q}^2$
is the four-momentum transfer. In the p(n,p)n reaction, however, this
is same as the three momentum transfer squared. $f_x$ is the $x$NN
coupling constant,
where $x$
denotes $\pi$ or $\rho$. $F_x(t)$ is the form factor at the $xNN$ vertex.
For it's form we use the monopole form, i.e.
\begin{equation}
F_x(t) = \frac {\Lambda ^2_x - m^2_x}{\Lambda ^2_x - t},
\end{equation}
where $\Lambda$ is the length parameter.

Substituting eq.(18) in eq.(16) the central and non-central parts
of the spin-isospin t-matrix, appearing in eqn. (7), work out as,
\begin{equation}
t_{\sigma \tau}^C(\mbox{\boldmath $q$}) = \int e^{i\mbox{\boldmath $q$} \cdot
\mbox{\boldmath $r$}} e^{i \xi (b)} V_{\sigma \tau}^C (r) d\mbox{\boldmath
$r$},
\end{equation}
and
\begin{equation}
t_{\sigma \tau}^{NC}(\mbox{\boldmath $q$})Y_{2M}(\hat q) =
\int e^{i\mbox{\boldmath $q$} \cdot \mbox{\boldmath $r$}} e^{i \xi (b)}
V_{\sigma \tau}^{NC}(r) Y_{2M}(\hat r) d\mbox{\boldmath $r$},
\end{equation}
where, in terms of the pi- and rho-exchange potentials,
in momentum space
\begin{equation}
V_{\sigma \tau}^C(t)=\frac {1}{3}[V_\pi (t)+2V_\rho (t)],
\end{equation}
and
\begin{equation}
V_{\sigma \tau}^{NC}(t)=\frac {1}{3}[V_\pi (t)-V_\rho (t)].
\end{equation}

A detailed presentation of the above t-matrix for the elementary
charge-exchange reaction and its applicability to the available
experimental data over a wide energy range is being reported
separately.

\section{Results and Discussions}
We calculate the double-differential cross-sections for the triton energy
spectrum at a
helium beam energy of 2 GeV and triton emission angles of $2^o$ to $7^o$.
Within the framework of the formalism given in the preceding sections,
various inputs, which determine these cross-sections, are :
(i) $\phi _{nl}$, the radial wave function of the neutron in the target
nucleus.
(ii) $^3$He$\rightarrow$t, transition form factor, $\rho (q)$.
(iii) Parameters associated with the one-pion- and rho-exchange
potentials.
(iv) Parameters of the elastic pn scattering amplitude, f(k,q), and
(v)  $^3$He and triton distorted waves.

The single particle wave functions, $\phi _{nl}$,
are generated in a Woods-Saxon potential, whose parameters are fixed from
the analyses of the  electron scattering and (p,2p) data on $^{12}C$ [15].
The neutron binding energies in the
$1s_{1/2}$ and $1p_{3/2}$ orbitals in $^{12}C$ are taken equal to 34 MeV
and 16 MeV respectively.

For the $^3$He$\rightarrow$t transition form factor, $\rho (q)$,
following the work of
Dmitriev et al. [16], we use,
\begin{equation}
\rho (q) = F_o\, e^{-\gamma q^2}[1 + \eta q^4],
\end{equation}
where $F_o = 0.96$, $\gamma $= 11.15 GeV$^{-2}$ and $\eta
$= 14 GeV$^{-4}$.
This form factor has been obtained using (s+d) wave mixed wave functions
for $^3$He and triton. The second term in the square brackets in
the above eqn.
is related to the d-wave admixture. This form factor is found to be
good upto large momentum transfers.

For the various parameters in the pion- and rho-exchange potentials,
we use $f_{\pi} = 1.008$,
$f_{\rho} = 7.815$, $\Lambda _{\pi}$ = 1.2 GeV/c and
$\Lambda _{\rho} $= 2 GeV/c. These values are consistent with several
experimental observations, like, $\pi$N scattering, NN scattering [17],
electro-disintegration of the deuteron [18], deuteron
properties [19] and dispersive theoretical approaches [20].
In the value of $f_{\rho}$, of course, there
is some uncertainty. It is 4.83, as determined by
the vector dominance model [21], and is 7.815 as determined from the
nucleon form factor and nuclear phenomena [22].

The elastic pn scattering amplitude, f(k,q), has three parameters:
the total cross-section, $\sigma _T$, the slope parameter, $\alpha $,
which determines it's momentum transfer (q) behaviour, and the
parameter, $\beta $, which determines the ratio of the real to imaginary
parts of f(k,q). Except $\beta $, both other parameters are well
known from the measured experimental data on the pn scattering [23].
The energy, k, at which these parameters need to be taken
is, of course, not defined well. This happens
 because of the Fermi motion of the nucleons in the
projectile and the target nuclei. However, as the beam energy
is high, for the
purpose of fixing the values of these parameters, we ignore the Fermi
motion of the proton in the beam and the struck neutron in the target.
The energy
of pn system in lab., therefore, is taken equal to 1/3 of the
$^3$He energy. Corresponding to 2 GeV $^3$He energy, we find
$\alpha \approx$ 6 (GeV/c)$^{-2}$ and $\sigma _T\approx $34
mb. In the value of $\beta $, there exists lot of uncertainty in
the `measured' values [24]. At the energy of our interest, it ranges
from 0.05 to -0.7. We have chosen the value -0.45 for our purpose.
The calculated cross-sections with this value of $\beta $ are found
in most reasonable agreement with the experimental data. We will,
of course, exhibit later the sensitivity of our results to the value of
$\beta $.

For the only iso-spin dependent part of the t-matrix, t$_\tau $,
we use the phenomenologically determined t-matrix of Franey and Love [11]
from NN scattering experiments at 725 MeV beam energy. The contribution
of this term to the cross-section, however, as we
shall see later, is not much.

The distorted waves for helium and triton appear as a product in eqs.
(10,11). In the eikonal approximation,
this can be approximated as follows,
\begin{equation}
\chi ^{-*}_t \chi ^+_{He} = e^{iq_{\parallel} z}\, e^{i
\mbox{\boldmath $q$}_\perp \cdot \mbox{\boldmath $b$}} e^{2i\delta (b)},
\end{equation}
if we ignore the difference
between
the phase shifts  of $^3$He and triton (around 2 GeV).
Here $\delta (b)$ corresponds to the phase shift
of a mass 3 particle. The
phase shift function $\delta (b)$ can be constructed, in principle, from
optical potentials. But, since this is a poorly known quantity,
we refer to the
experimentally measured values. The experimental $\delta (b)$ too is not
available for mass 3 particles and hence we use the phase shifts obtained
from alpha scattering at
1.37 GeV on calcium isotopes [25]. Here, $exp[2i\delta (b)]$, which
gives a good description of the $\alpha$ scattering data, is found to be
purely real and has a 1 minus Woods-Saxon form. The radius parameter
$r_o(R=r_oA^{1/3})$, and diffuseness, a, of this functional form are found
equal to 1.45 and 0.68 fm. respectively. We use this phase shift
function for our purpose too, except that the radius, R, is put
corresponding to A=12.

Before we present the calculated cross-sections, in fig. 2 we show
at 2 GeV beam energy the range of momentum transfer (q) involved
in the triton emission
upto 7$^0$ in lab. In the quasi-elastic range
of the triton energy (i.e. upto 1900 MeV),
this momentum transfer, as we see, is not small. At 2-3 degree it is around
250 MeV/c, and at larger angles it goes to about 500 MeV/c. This
suggests that the non-central
component of the t-matrix, T$_{BA}$, and hence the $\rho $-exchange
part of the $V_{\sigma \tau}$ (whose contribution increases at larger q),
may effect  the
quasi-elastic cross-section significantly. However, because
the rho-exchange
contributes with opposite signs to the central and non-central pieces
of the potential (see eqs. 23, 24), it is not immediately obvious as
how much, in net, the rho-exchange would change the cross-sections.
In fig. 3 we plot the typical central and non-central components
of the real part of the calculated elementary t-matrix used in
our calculations. We show this t-matrix for a pure one-pion-exchange
potential and for a one pion- plus rho-exchange potential.
As expected, we see that the rho-exchange affects both the pieces
of the t-matrix significantly,
but in the opposite directions.

To demonstrate the extent to which
the above t-matrix
reproduces the measured p(n,p)n cross-sections, in fig. 4 we show
the calculated 0$^0$ cross-sections  alongwith the experimental data
[26-33] over a large energy range. As
we see, the calculated cross-sections are in good accord with the
measured cross-sections.

In fig. 5 we show the results for the $^{12}$C($^3$He,t) reaction together
at all the
angles of the triton emission. In fig. 5a we see a representation of the
experimentally measured cross-sections and in fig. 5b that of our
corresponding theoretically
calculated results. The calculated cross-sections are obtained with
the pion and rho meson both included, and with contributions from the
spin-flip
and non-spin-flip channels to the transition matrix. As can be seen from the
figs., the overall behaviour of the experimental cross-sections, which
includes the magnitude and position of the peak and its shift with the
emission angle of the triton, gets reproduced reasonably well by
the theoretical calculations.  This vindicates, in essence,
the applicability of the quasi-free mechanism framework presented
in the earlier section to the region of the triton spectrum
lying between the bound nuclear states and the delta production
region.

In the following, we give the results at each angle separately.

In figs. 6, 7 and 8 we show the triton energy spectra plotted individually for
six triton emission angles between $2^o$ and $7^o$. The solid curves
represent the
calculations with the one-pion plus one-rho exchange interaction. The
experimental results are represented by the dots. Except at $5^o$ and
$7^o$, we find a good accord between the calculated and measured cross-
sections. The underestimation of the cross-sections at  $7^o$ should
not be a source of much discouragement,
as the magnitude of the cross-section is too small
at this angle ($\sim  7\mu b$). Therefore, the measured cross-section
can have large
uncertainty and a significant
contribution from other reaction mechanisms. The reason for the
overestimation of the cross-section by about a factor of 3/2
at 5$^0$ is, of course, not clear to us.

To isolate the contribution due to rho-exchange to the calculated
cross-sections, in figs. 6-8 we also display, by dashed curves, the
cross-sections
with only the one-pion-exchange transition potential. As
we see from these figs., the contribution of the rho-exchange changes
continuously from being
positive at $2^o$ to negative at $7^o$. Around 3$^0$ it crosses the
zero level. This happens, as mentioned earlier, due to change in
the momentum transfer, q, (see fig. 2) and the opposite signs of the
rho-exchange potential in the central and the non-central pieces of the
potential. At smaller momenta, where the central term dominates, the
rho-exchange comes with the positive sign, while at larger angles,
where the non-central term is important, it comes with the negative
sign (see eqs. 23, 24 and fig. 3).

In our calculations, we also find that the calculated cross-sections are
mainly decided by the spin-isospin dependent part of the t-matrix. This
can be
seen in fig. 9, where we show the 2$^0$ calculated cross-sections
with and without the isospin-dependent
term, $t_{\tau}$. The dashed curve  represents the calculation without
the
isospin-dependent term in the t-matrix, while the solid curve is the
complete
calculation. As we see, the only isospin-dependent term makes a
contribution of less than
$10\%$ to the cross-section.

As we mentioned earlier, the only uncertain parameter in the above
calculations had been the value of the parameter $\beta $ - the
ratio of the real to imaginary part of the elementary scattering
amplitude. In fig. 10 we show the calculated triton energy spectrum
at 2$^0$ and 6$^0$ for three values of $\beta $, viz. 0.05, -0.45 and
-0.7.
These values lie within the uncertainty of  the experimentally
extracted value.
As we see, the calculated cross-sections do depend upon the value of
$\beta $.  For the present range, it can change the peak cross-section
by a factor of 2.

The quasi-free cross-sections, as we see from the examination of the
expression (eqs. 6,7) for $\langle |T_{BA}|^2 \rangle$, is
essentially determined by, (i)
the t-matrix, t$_{\sigma \tau} $, and (ii) the neutron momentum
distribution
in the target nucleus through the recoil momentum distribution factors,
$F^{\nu ,+1}(\mbox{\bf Q})$ and $F^{+1}(\mbox{\bf Q})$.
The t$_{\sigma \tau} $ depends upon q, and F's on Q.
In fig. 2 we see  that the magnitude of q,
at a particular triton
emission angle, does not vary much in the region of the quasi-elastic peak.
This means that
the elementary t-matrix, t$_{\sigma \tau} $,
too does not change much over this region for a
fixed triton emission angle. Consequently, in the A($^3$He,t)B reaction,
the elementary t-matrix mainly affects the
magnitude of the cross-sections (see e.g. fig. 10).
The shapes of the triton energy spectra, which in the inclusive
data mean the peak position and the width, are decided by the
the recoil momentum distribution factors. Since in the inclusive
data the proton in the final state is not detected, and
$\mbox{\bf Q} =\mbox{\bf k}_{He} - \mbox{\bf k}_t -
\mbox{\bf k}_p $, each point in the triton energy spectrum involves an
integral over a certain range of Q.
This means that even the shape of the triton energy spectra might
not depend upon the details of the neutron momentum distribution
in the nucleus. It may be
sufficient if the  neutron wave functions have correct
separation energies for different shells and reproduce some
gross properties, like the r.m.s. radius, of the nucleus. To exhibit
this, in fig. 11 we show the calculated cross-sections
for two radial wave functions, $\phi _{nl}$, of the neutron in
the target nucleus.  The solid curve is the calculation
with $\phi _{nl}$ generated in
a Woods-Saxon potential as has been used through out this work; the dashed
curve is the calculation using harmonic oscillator wave function with the
oscillator parameter b=1.66 fm, which is consistent with the electron
scattering data. As can be seen from the figure,
within about $10\%$, the two results are same.

\section{Conclusions}

We have examined the quasi-elastic peak region in the $^{12}$C($^3$He,t)
reaction
at 2 GeV over a range of triton emission angles from $2^o$ to $7^o$.
We have
calculated the triton energy spectra in the framework of a
quasi-elastic charge-exchange between a proton in the
projectile and a neutron in the target
nucleus.  Constructing the t-matrix for this process with the
$\pi + \rho$ transition potential, and using distorted waves for $^3$He
and triton, the overall features of the experimentally measured
cross-sections are produced reasonably well. Various inputs used for
these calculations are constrained by other known experimental
quantities, and thus are not arbitrary.

\section{\bf Acknowledgements}
One of the authors (N.G.K.) wishes to thank the support given by the
Department
of Atomic Energy, Government of India, for part of the work
which was
done at the Bhabha Atomic Research Centre in Bombay.

\newpage
{\large\bf Figure Captions}
\vskip 1cm
\begin{itemize}
\item[1.]
Diagrammatic representation of the reaction
$^3He + \, ^{12}C \rightarrow t + p + \, ^{11}C$.
\item[2.]
Kinematics for the $^{12}C(^3He,t)$ reaction at T$_{He}$ = 2 GeV. The triton
energies are plotted as a function of the momentum transfer q, for triton
emission angles of $0^o$ to $7^o$.
\item[3.]
Real part of the t-matrix as a function of the momentum transfer $\mbox{\bf q}
= \mbox{\bf k}_{He} - \mbox{\bf k}_t$. Solid curves
represent the t-matrix constructed with the $\pi + \rho$ exchange transition
potential and dashed curves the one-pion exchange only. (a) Central part of the
t-matrix (b) Non-central part of the t-matrix.
\item[4.]
Calculated and measured p(n,p)n cross-sections at 0$^0$ for various
neutron energies. The calculated cross-sections are for one pion-
plus rho-exchange spin-isospin potential.
\item[5.]
Triton energy spectra in the quasi-elastic peak region for $^{12}C(^3He,t)$
at incident beam energy T$_{He}$ = 2 GeV and triton emission angles of $2^o$ to
$7^o$. (a) Representation of the experimental data from ref.[3]. (b)
Theoretical calculations with the $\pi + \rho$ exchange transition potential.
\item[6.]
Triton energy spectra in the quasielastic peak region at triton angles of
$2^o$ and $3^o$ for $^{12}C(^3He,t)$ at 2 GeV beam energy. The solid curves are
the cross-sections with the t-matrix consisting of the $\pi + \rho$ exchange
transition potential. The dashed curves are the results with only one-pion
exchange. Dots represent the experimental data (ref.[3]).
\item[7.]
Same as fig. 3 for triton emission angles of $4^o$ and $5^o$.
\item[8.]
Same as fig. 3 for triton emission angles of $6^o$ and $7^o$.
\item[9.]
Comparison of the contribution from the spin-isospin-flip channel (solid
curve) and non-spin-flip channel (dashed curve) to the triton energy spectrum
at $2^o$ for the $^{12}C(^3He,t)$ reaction at 2 GeV. Dots represent the
experimental data (ref.[3]).
\item[10.]
Triton energy spectra in the quasielastic peak region at triton angles of
$2^o$ and $6^o$ for $^{12}$C($^3$He,t) at 2 GeV beam energy for
different values of $\beta $. The curves are
the cross-sections with the t-matrix consisting of the $\pi + \rho$ exchange
transition potential. The numbers on the curves indicate the corresponding
values of $\beta $.
\item[11.]
Triton energy spectrum at triton angle of $2^o$ and 2 GeV beam energy with
different forms of the radial wave function $\phi _{nl}$ of the neutron in
the target nucleus.
\end{itemize}

\newpage
\begin{thebibliography}{26}
\bibitem{chr} G. Jacob and A. J. Maris,   Nucl. Phys.
{\bf 31}, 139 (1962); {\it ibid}
Rev. of Mod. Phys. {\bf 38}, 121 (1966);  T. Berggren and H. Tyren,
Ann. Rev. Nucl. Sci. {\bf 16}
, 153 (1966); D. F. Jackson, Adv. Nucl. Phys. {\bf 4}, 1 (1971);
R. E. Chrien {\it et al.}, Phys. Rev. {\bf C21}, 1014 (1980);
J. S. O'Connell {\it et al.}, Phy. Rev. {\bf C35}, 1063 (1987).
\bibitem{bon}B. E. Bonner {\it et al.}, Phy. Rev. {\bf C18}, 1418 (1978);
C. Gaarde, Nucl. Phys. {\bf A507}, 79c (1990).
\bibitem{berg}I. Bergqvist {\it et al.}, Nucl. Phys. {\bf A469}, 648 (1987);
\bibitem{brow}G. E. Brown, J. Speth and J. Wambach, Phy. Rev. Lett. {\bf 46}
, 1057 (1981).
\bibitem{th} W. M. Alberico, M. Ericson and A. Molinari, Phys. Lett.
{\bf 92B}, 153 (1980); {\it ibid} Nucl. Phys. {\bf A379}, 429 (1982);
\bibitem{cont}D. M. Contardo {\it et al.}, Phys. Lett. {\bf 168B}, 331 (1986);
C. Ellegard {\it et al.}, Phys. Rev. Lett. {\bf 50}, 1745 (1983); {\it ibid}
Phys. Lett. {\bf 154B}, 110 (1985); D. Bachelier {\it et al.}, Phys. Lett.
{\bf 172B}, 23 (1986).
\bibitem{alber}
W. M. Alberico {\it et al.}, Phys. Rev. {\bf C38}, 109 (1988).
\bibitem{ichi}
M. Ichimura {\it et al.}, Phys. Rev. {\bf C39}, 1446 (1989).
\bibitem{bert}
G. F. Bertsch, L. Frankfurt and M. Strkman, Science {\bf 259}, 773 (1993).
\bibitem{jaisan}
B. K. Jain and A. B. Santra, Phys. Rev. {\bf C46}, 1183 (1992);
{\it ibid} Phys. Rep. {\bf 230}, 1 (1993).
\bibitem{love}
W. G. Love and M. A. Franey, Phy. Rev. {\bf C24}, 1073 (1981);
M. A. Franey and W. G. Love, Phys. Rev. {\bf C31}, 488 (1985);
\bibitem{aust} N. Austern and C. M. Vincent, Phys. Rev. {\bf C23},
1847 (1981).
\bibitem{abs}
B. K. Jain and A. B. Santra, Nucl. Phys. {\bf A519}, 697 (1990);
{\it ibid} Phy. Lett. {\bf B244}, 5 (1990).
\bibitem{pale}
H. Palevsky {\it et al.}, Phys. Rev. Lett. {\bf 18}, 1200 (1967); G. J. Igo
{\it et al.}, Nucl. Phys. {\bf B3}, 181 (1967); B. A. Ryan {\it et al.},
Phys. Rev. {\bf D3}, 1 (1971).
\bibitem{elt}
L. R. B. Elton and A. Swift, Nucl. Phys. {\bf A94}, 52 (1967);
R. Shanta and B. K. Jain, Nucl. Phys. {\bf A175}, 417 (1971).
\bibitem{dmit}
V. F. Dmitriev, O. Sushkov, and C. Gaarde, Nucl. Phys.
{\bf A459}, 503 (1986).
\bibitem{mach}
R. Machleidt, K. Holinde and C. Elester, Phys. Rep. {\bf 149}, 1 (1987); W.
N. Cottingham {\it et al.}, Phys. Rev. {\bf D8}, 800 (1973).
\bibitem{math}
J. F. Mathiot, Nucl. Phys. {\bf A412}, 201 (1984).
\bibitem{eric}
T. E. O. Ericson and M. Rosa-Clot, Nucl. Phys. {\bf A405}, 497 (1983).
\bibitem{nutt}
W. Nutt and B. Loiseau, Nucl. Phys. {\bf B104}, 333 (1976).
\bibitem{nage}
M. M. Nagels {\it et al.}, Nucl. Phys. {\bf B109}, 1 (1976).
\bibitem{hohl}
G. Hohler and E. Pietarinen, Nucl. Phys. {\bf B95}, 210 (1975).
\bibitem{silv}
B. H. Silverman {\it et al.}, Nucl. Phys. {\bf A499}, 763 (1989).
\bibitem{grei}
W. Grein, Nucl. Phys. {\bf B131}, 255 (1977).
\bibitem{chou}
D. C. Choudhury, Phys. Rev. {\bf C22}, 1848 (1980).
\bibitem{shep}
P. F. Shepard, T. J. Devlin, R. E. Mischke and J. Solomon, Phys. Rev.
{\bf D10}, 2735 (1974); R. E. Mischke, P.F. Shepard and T.J. Devlin, Phy. Rev.
Lett. {\bf 23}, 542 (1969).
\bibitem{bonn}
B. E. Bonner {\it et al.}, Phy. Rev. Lett. {\bf 41}, 1200 (1978).
\bibitem{hurs}
W. Hurster {\it et al.}, Phys. Lett. {\bf B90}, 367 (1980).
\bibitem{biz}
G. Bizard {\it et al.}, Nucl. Phys. {\bf B85}, 14 (1975).
\bibitem{kaza}
Yu. M. Kazarinov and Yu. N. Simonov, Zh. Eskp. Teor. Fiz. {\bf 43}, 35 (1962)
(Sov. Phys. JETP {\bf 16}, 24 (1963)).
\bibitem{how}
V. J. Howard {\it et al.}, Nucl. Phys. {\bf A218}, 140 (1974).
\bibitem{meas}
D. F. Measday, Phy. Rev. {\bf 142}, 584 (1966).
\bibitem{palmi}
J. N. Palmieri and J.P. Wolfe, Phy. Rev {\bf C3}, 144 (1971).
\end{thebibliography}
\end{document}